\documentclass{IEEEtran}
\usepackage{cite}
\usepackage{amsmath,amssymb,amsfonts}
\usepackage{algorithmic}
\usepackage{graphicx}
\usepackage{bm}
\usepackage{textcomp}
\usepackage{fancyhdr}
\usepackage{booktabs}
\usepackage{color}
\usepackage[table,xcdraw]{xcolor}
\usepackage{subfigure}
\def\BibTeX{{\rm B\kern-.05em{\sc i\kern-.025em b}\kern-.08em
    T\kern-.1667em\lower.7ex\hbox{E}\kern-.125emX}}

\usepackage{soul}
\usepackage{makecell}
\usepackage[hyphens]{url}

\fancypagestyle{plain}{
	\fancyhf{} 
	\fancyfoot[C]{\textbf{\thepage}} 
	
	}

\makeatletter
\long\def\@makecaption#1#2{
	\vskip\abovecaptionskip
	\sbox\@tempboxa{#1. #2}
	\ifdim \wd\@tempboxa >\hsize
	#1. #2\par
	\else
	\global \@minipagefalse
	\hb@xt@\hsize{\hfil\box\@tempboxa\hfil}
	\fi
	\vskip\belowcaptionskip}
\makeatother

\makeatletter
\def\thickhline{%
	\noalign{\ifnum0=`}\fi\hrule \@height \thickarrayrulewidth \futurelet
	\reserved@a\@xthickhline}
\def\@xthickhline{\ifx\reserved@a\thickhline
	\vskip\doublerulesep
	\vskip-\thickarrayrulewidth
	\fi
	\ifnum0=`{\fi}}
\makeatother

\newlength{\thickarrayrulewidth}
\usepackage{caption}
\begin{document}

\onecolumn
\newpage
\thispagestyle{empty}

\textbf{Copyright:}\\
\copyright 2020 IEEE. Personal use of this material is permitted.  Permission from IEEE must be obtained for all other uses, in any current or future media, including reprinting/republishing this material for advertising or promotional purposes, creating new collective works, for resale or redistribution to servers or lists, or reuse of any copyrighted component of this work in other works.\\

\textbf{Disclaimer:}\\
This work has been published on \textit{ IEEE Transactions on Circuits and Systems II: Express Briefs}. DOI: 10.1109/TCSII.2020.3000058\\

\newpage

\title{
Low-Loss Reconfigurable Phase Shifter in Gap-Waveguide Technology for mm-Wave Applications}
\author{\'{A}ngel Palomares-Caballero, Antonio Alex-Amor, Pablo Escobedo, Juan Valenzuela-Vald\'{e}s, Pablo Padilla
\thanks{This  work  was  supported  in  part  by  the  Spanish Research and Development National Program under Project TIN2016-75097-P and Project RTI2018-102002-A-I00, in part by “Junta de Andalucia” under Project  B-TIC-402-UGR18  and  Project  P18.RT.4830, and in part by the Predoctoral Grant FPU18/01965. \textit{(Corresponding author: \'{A}ngel Palomares-Caballero.)}}

\thanks{\'{A}. Palomares-Caballero, A. Alex-Amor, Juan Valenzuela-Vald\'{e}s and P. Padilla are with the Departamento de Teor\'{i}a de la Se\~{n}al, Telem\'{a}tica y Comunicaciones, Universidad de Granada, 18071 Granada, Spain (e-mail: angelpc@ugr.es; aalex@ugr.es; juanvalenzuela@ugr.es; pablopadilla@ugr.es).}
\thanks{A. Alex-Amor is with the Information Processing and Telecommunications Center, Universidad Polit\'{e}cnica de Madrid, 28040 Madrid, Spain (e-mail: aalex@gr.ssr.upm.es).}
\thanks{P. Escobedo is with Bendable Electronics and Sensing Technologies (BEST) Group, School of Engineering, University of Glasgow, Glasgow G12 8QQ, UK (e-mail: pablo.escobedo@glasgow.ac.uk).}
}

\twocolumn

\maketitle

\begin{abstract}
In this brief, we present a low-loss mechanically reconfigurable phase shifter implemented in gap-waveguide technology for mm-wave frequencies. The proposed design gives a practical implementation of tuning elements inside the waveguide providing alternatives to the use of the E-plane split waveguide at high frequencies in order to avoid leakage losses. The depicted phase shifter design is based on a H-plane split waveguide. The phase shift is controlled by means of a tuning screw, which exerts pressure on a flexible metallic strip inserted inside the waveguide. The flexible strip bends with different curvature radii and determines the phase shift at the output port. Cost-effective manufacturing and simple implementation of the flexible metallic strip are achieved by means of the gap-waveguide design. A prototype has been manufactured for validation purposes. Good impedance matching is achieved from 64 GHz to 75 GHz providing a 15.8\% impedance bandwidth. The results show a maximum phase shift of 250\textsuperscript{o} with a maximum and mean insertion loss (IL) of 3 dB and 1.7 dB, respectively.

\end{abstract}

\begin{IEEEkeywords}
Phase shifter, gap-waveguide technology, low-loss, reconfigurable waveguide, mm-waves.
\end{IEEEkeywords}

\section{Introduction}
\IEEEPARstart{R}{econfigurability} is playing a fundamental role in present and future mm-wave communication systems.  Thus, reconfigurable phase shifters are necessary to provide beamforming capabilities to the antennas  \cite{MultibeamAntennas}. However, this kind of phase shifters has to be implemented in a proper technology to present low losses and fulfill demanding requirements. Waveguide technology is a good candidate since it provides reduced losses in the mm-wave frequency range. Some examples of mechanically reconfigurable phase shifters in waveguide technology at K-band can be found in \cite{MechanicalPhaseShifter1,MechanicalPhaseShifter2}. In both designs, short waveguide couplers with mechanical reconfigurable elements at the output ports have been used to achieve the desired phase shift in reflection. Another solution in waveguide technology is presented in \cite{MechanicalPhaseShifter3}, where the wide side of the waveguide is adjustable to achieve the desired phase shift at the output port of the waveguide. Similarly, \cite{PiezoelectricPhaseShifter} shows a reconfigurable phase shifter based on a piezoelectric actuator, where the side wall of the waveguide is in this case a perfect magnetic conductor (PMC). Other approaches to obtain phase shift employ PIN diodes \cite{PinDiodePhaseShifter0,PinDiodePhaseShifter1,PinDiodePhaseShifter1_2,PinDiodePhaseShifter2,PinDiodePhaseShifter3,PinDiodePhaseShifter3_2,PinDiodePhaseShifter4}. Nevertheless, PIN diodes suffer from high losses in mm-wave regime. Some reconfigurable material such as liquid crystals \cite{LcPhaseShifter1,LcPhaseShifter2,LcPhaseShifter3,LcPhaseShifter4,LcPhaseShifter5} or ferroelectrics \cite{Ferroelectric1,Ferroelectric2,Ferroelectric3} are also used to vary the phase shift. These approaches are currently attractive but present some drawbacks such as complex implementation in waveguide and high insertion losses incurred when the frequency increases. 

\begin{figure}[t]
	\centering
	\includegraphics[width=0.42 \textwidth]{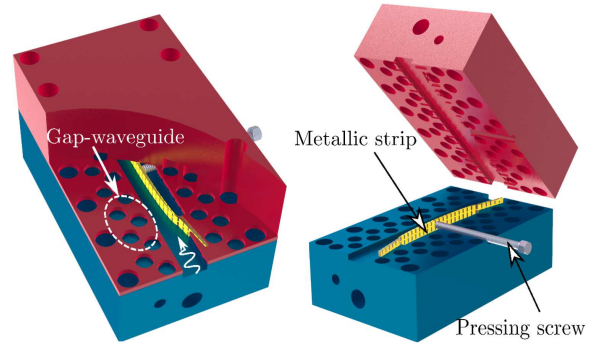}
	\caption{Tunable phase shifter design model.} 
	\label{Figure1}
\end{figure}

Waveguide technology has evolved in order to provide a less demanding manufacturing process. Gap-waveguide technology \cite{GapWave1} enables the waveguide fabrication in two separate parts with no mandatory perfect electrical contact in the assembly, avoiding possible leakage throughout the gap. This fact permits to relax the requirements of manufacturing providing a low-cost fabrication. Besides, taking advantage of the waveguide fabrication in splitted parts, tuning elements can be easily implemented inside the waveguide in order to modify the phase of the propagating signal. For example, in \cite{GapWavePhaseShifter1} and \cite{GapWavePhaseShifter2}, gap-waveguide based on glide-symmetric holes \cite{GapWave2} has been used to design a mm-wave phase shifter inserting a dielectric slab and pinned structures, respectively. This work presents a tunable phase shifter based on gap-waveguide technology. The reconfigurable behavior of the phase shifter is achieved by means of a flexible metallic strip inserted in the side wall of the waveguide. The variation of the curvature radius of the flexible metallic strip is produced by a tuning screw that exerts a pressure on the center of the strip. It allows to narrow the wide side of the waveguide and modify the propagation constant ($\beta_{g}$), producing an adjustable phase shift. Figure \ref{Figure1} depicts the design of the proposed tunable phase shifter.

Some patents about mechanically tunable phase shifter in conventional waveguide technology are in \cite{Patent1,Patent2}. The latter inserts a pair of dielectric rods inside the waveguide in order to change the phase shift. However, low phase shift is achieved and increase in the insertion losses are expected due to the wave propagation in the dielectric material at high frequencies. Additionally, no in-depth details are provided about the manufacturing issues and performance of a real prototype. There exist  examples of commercially available phase shifters in waveguide technology \cite{PasternackPhaseShifter,SagePhaseShifter}. Nevertheless, their way of implementing the phase shift in waveguide is not reported. In addition, they are bulky and high-cost designs due to the complexity of implementing the phase shift mechanism. The proposed phase shifter design exploits the advantages of the gap-waveguide technology to allow a cost-effective device with low losses. To the best of the authors' knowledge, this is the first reported mechanically reconfigurable gap-waveguide phase shifter in mm-waves frequencies. 

The paper is organized as follows. Section II presents the phase shifter design and its reconfigurable behavior. Section III shows the experimental results obtained with the prototype as well as the comparison with the simulation results. Finally, conclusions are provided in Section IV.

\section{Phase Shifter Design}

\begin{figure}[t]
	\includegraphics[width=0.44 \textwidth]{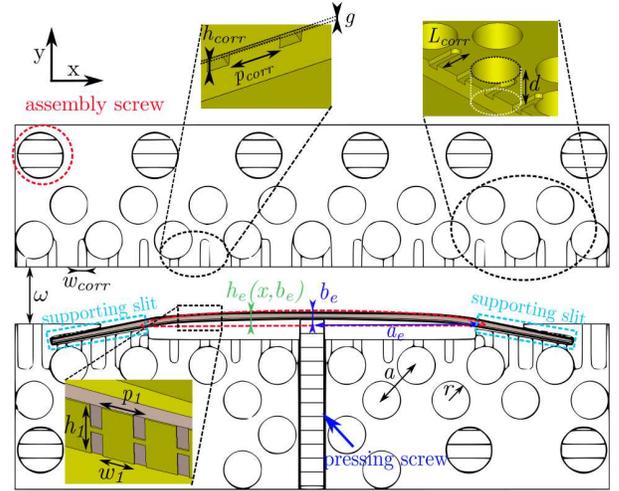}
	\centering
	\captionsetup{labelfont={color=black}}
	
	\caption{Description of the upper layer of the phase shifter. Dimensions: \textit{a\textsubscript{e}} = 11 mm, \textit{b\textsubscript{e}} = 0.55 mm, $\omega$ = 3.76 mm, \textit{a} = 3.22 mm, \textit{r} = 1.25 mm, \textit{w\textsubscript{corr}} = 0.6 mm, \textit{h\textsubscript{corr}} = 0.25 mm, \textit{p\textsubscript{corr}} = 2 mm, \textit{L\textsubscript{corr}} = 1.8 mm, \textit{h\textsubscript{1}} = 1.8 mm, \textit{p\textsubscript{1}} = 1.1 mm, \textit{w\textsubscript{1}} = 0.8 mm.}
	\label{Figure2}
\end{figure}

A top view of the proposed phase shifter is illustrated in Fig. \ref{Figure2}. The contour of the metallic strip can be modeled with an elliptical profile, whose semi-major axis \textit{a\textsubscript{e}} is fixed and whose semi-minor axis \textit{b\textsubscript{e}} depends on the position of the tuning screw. The more pressure the screw exerts, the larger \textit{b\textsubscript{e}} is. Since the waveguide works in its fundamental mode, the propagation constant $\beta_g$ for the TE\textsubscript{10} \cite{Pozar} is presented in Eq. (1). The right-hand side of this expression shows the dependence with the narrowing of the waveguide ($\omega-h_{e}(x,b_e)$). Therefore, when the signal propagates along the waveguide, the propagation constant $\beta_{g}(x,b_e)$ varies according to selected position $x$ of the waveguide and \textit{b\textsubscript{e}}. Note that fundamental mode is considered in the entire length of the phase shifter, 2\textit{a\textsubscript{e}}. The expressions for $\beta_{g}(x,b_e)$ and the approximated phase shift $\phi$ obtained along the waveguide are represented as:

\begin{equation}\label{beta_g} 
\resizebox{.9\hsize}{!}{$\beta_{g}(x,b_e) = \dfrac{2\pi}{\lambda_{g}(x,b_e)} = 2\pi\dfrac{\sqrt{1-\left( \dfrac{\lambda_{o}}{2(\omega-h_{e}(x,b_e))}\right)^2}}{\lambda_{o}}$}
\end{equation}

\begin{equation}\label{equationPhase} 
\phi(b_e) = \int_{-a_{e}}^{a_{e}} \beta_{g}(x,b_e) \, dx
\end{equation}

\noindent where $\lambda_{o}$ is the free-space wavelength and $w$ is the size of the broadside in waveguide rectangular size 15 (WR15). From Eqs. (1)-(2), it is clear that a variation in the profile of the ellipse produces a modification in the total phase shift. The selection of the dimensions \textit{a\textsubscript{e}} and \textit{b\textsubscript{e}} are based on the proposed model and represent a tradeoff between three main aspects of the phase shifter: operational frequency range, maximum phase shift and compactness. The phase shift results provided by this model are shown at the end of this section compared with the simulations in \textit{CST Microwave Studio}.

\begin{figure}[t]
	\centering
	\captionsetup{labelfont={color=black}}
	
	\subfigure[]{\includegraphics[width= 0.255\textwidth]{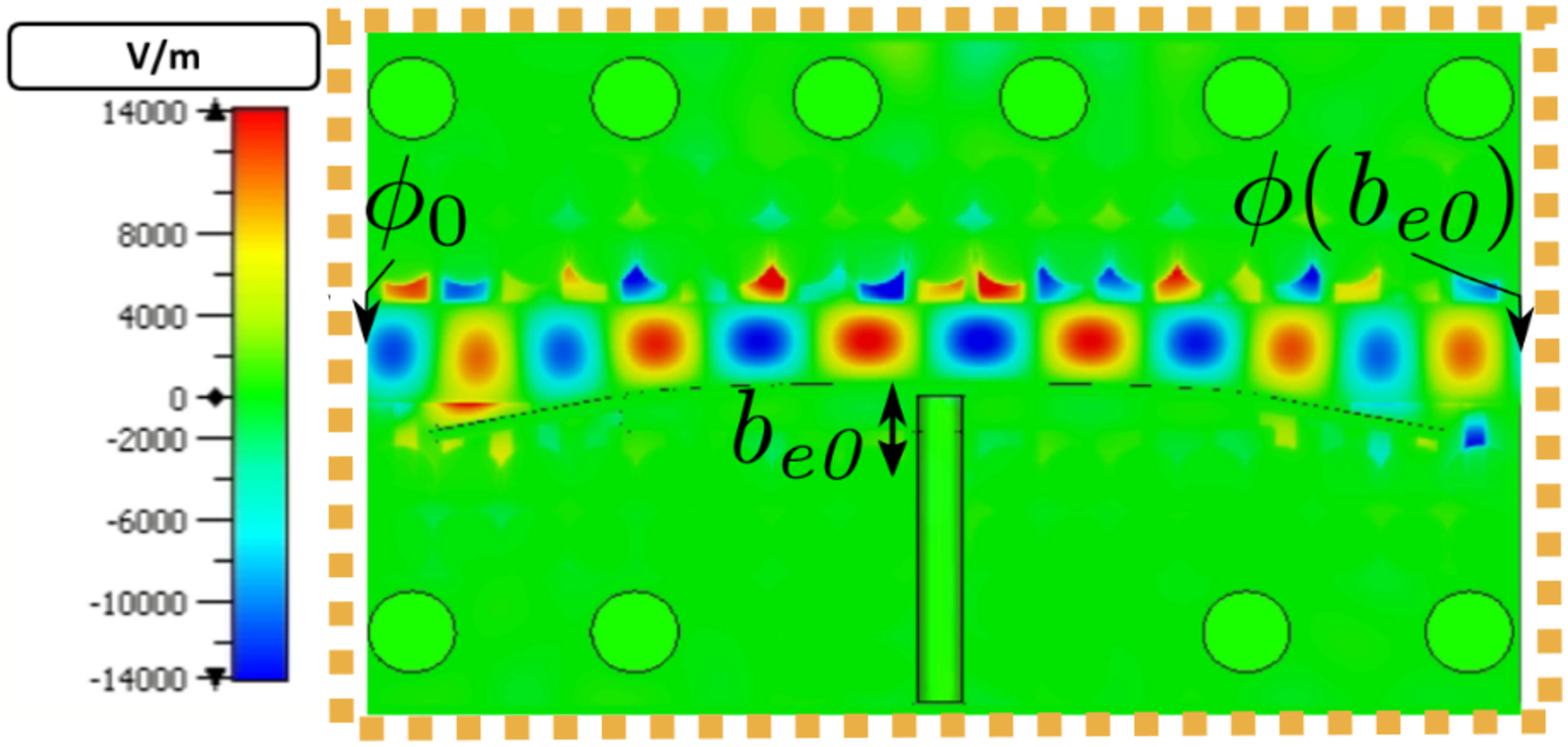}\label{Figure3a}}
	\hspace{0mm}
	\subfigure[]{\includegraphics[width= 0.205\textwidth]{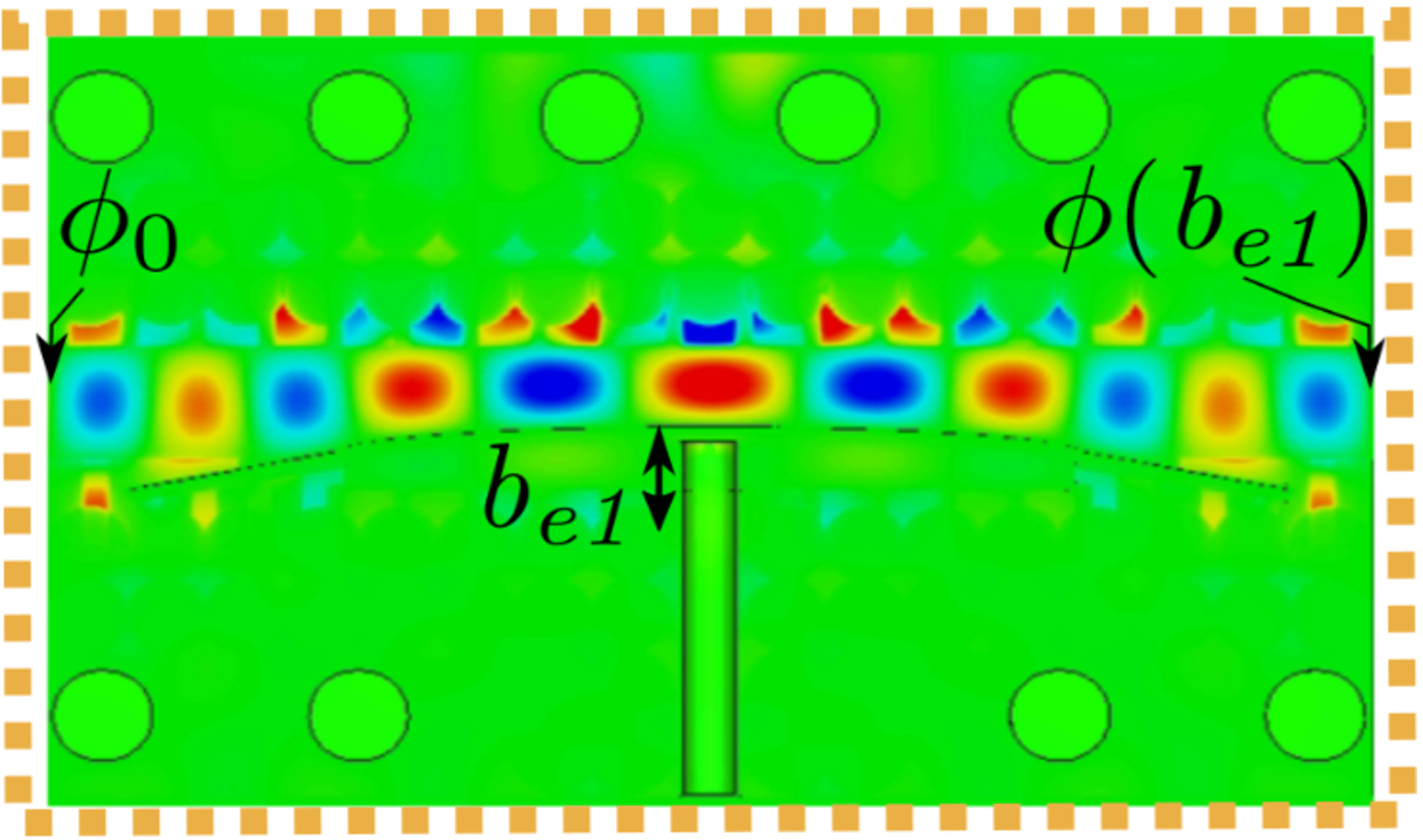}\label{Figure3b}}
	\hspace{0mm}
	\subfigure[]{\includegraphics[width= 0.42\textwidth]{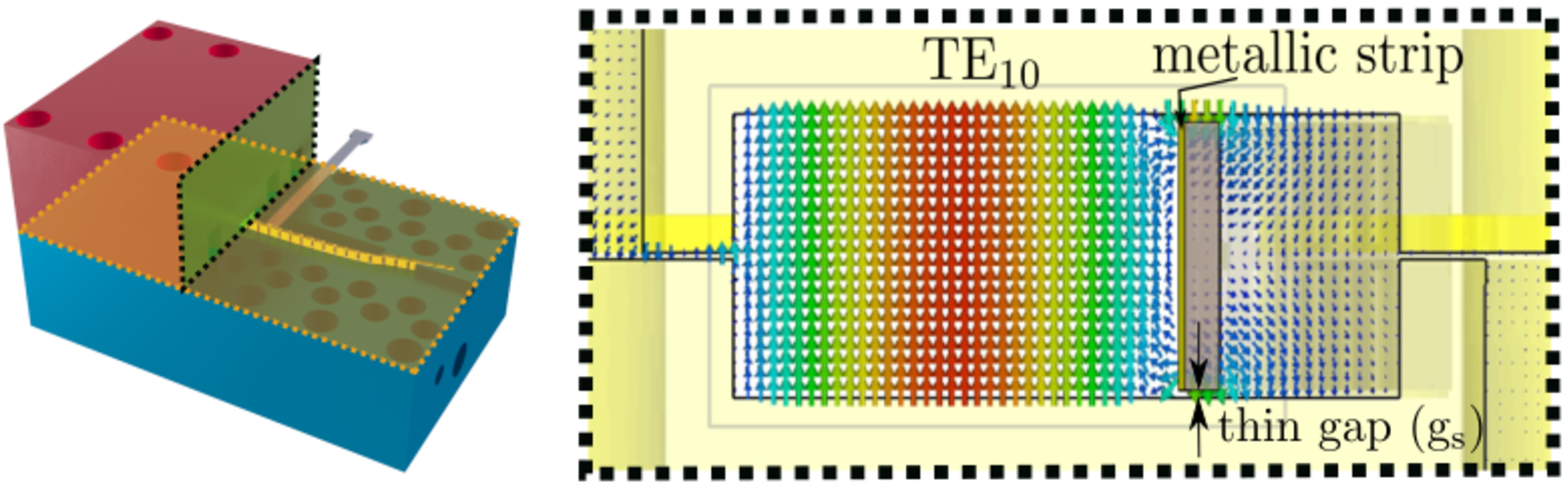}\label{Figure3c}}
	
	\caption{Amplitude of the E-field distribution: (a) top view of the gap between layers at 65 GHz when the tuning screw does not exert any pressure, (b) when the tuning screw exerts a pressure. The phase changes as a consequence of the tuning screw. (c) 3D views of the selected cutting planes and transversal cut of the phase shifter showing the fundamental propagative mode.}
	\label{Figure3}
\end{figure}

In order to allow an easy implementation of the metallic strip inside the waveguide, gap-waveguide technology based on glide-symmetric holes is chosen \cite{GapWave2}. The dimensions for the glide-symmetric holes are depicted in Fig. \ref{Figure2}, where the gap height \textit{g} and the depth of the holes \textit{d} are set at 0.05 mm and 3 mm, respectively. This electromagnetic bandgap (EBG) structure provides a stopband for the gap leakage from 45 GHz to 85 GHz, covering the frequency range of the waveguide standard WR15. Figures \ref{Figure3a} and \ref{Figure3b} show the effect of the gap-waveguide technology in the proposed phase shifter. It is observed how the phase shift can be controlled by changing the curvature radius of the metallic strip. 

\begin{figure}[t]
	\centering
	\captionsetup{labelfont={color=black}}
	
	\subfigure[]{\includegraphics[width= 0.49\textwidth]{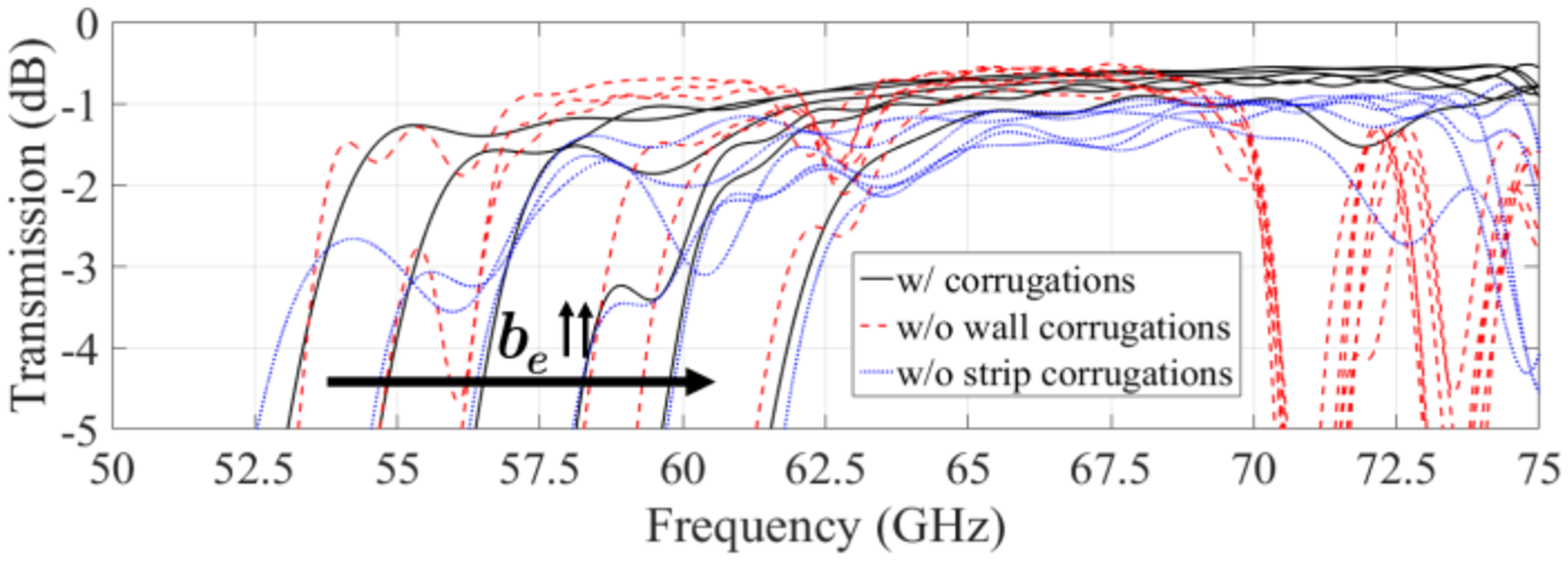}\label{Figure4a}}
	\hspace{0mm}
	\subfigure[]{\includegraphics[width= 0.49\textwidth]{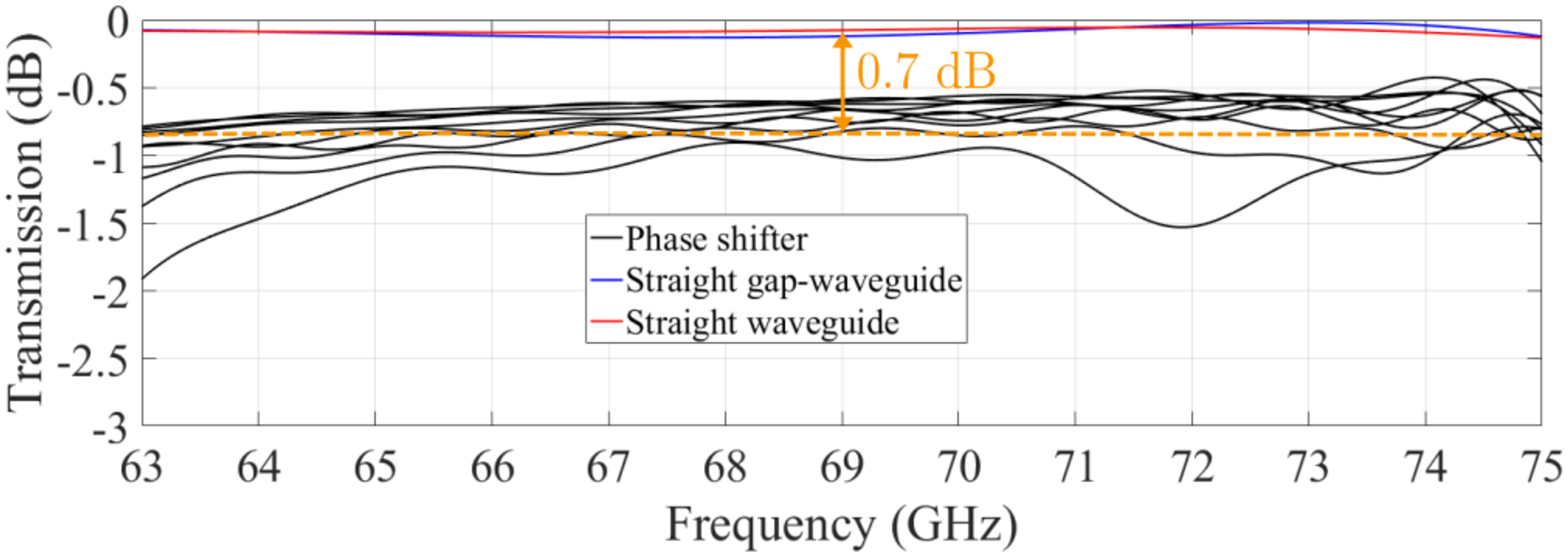}\label{Figure4b}}
	\hspace{0mm}

	\caption{Simulated transmission coefficient: (a) with (w/) and without (w/o) strip and wall corrugations for different curvature radii (different \textit{b\textsubscript{e}} values). (b) Comparison with a straight gap-waveguide and conventional waveguide that have the same length as the phase shifter (for different curvature radii).}
	\label{Figure4}
\end{figure}

In order to prevent undesired resonances between glide-symmetric EBG holes and the waveguide, corrugations are implemented to avoid resonant fields \cite{Corrugations},\cite{Corrugations2}. 
In addition, a vertical thin gap g\textsubscript{s} (shown in Fig. \ref{Figure3c}) should exist between the long side of the metallic strip and the upper and lower surface of the waveguide to permit the flexible movement of the metallic strip. These thin gaps may cause resonant fields in the same manner as previous situation. For that reason, the flexible metallic strip has also a corrugated shape in order to prevent resonances. A transversal cut of the phase shifter is illustrated in Fig. \ref{Figure3c}, where 50 $\mu$m gaps are included in both sides of the metallic strip to observe the E-field distribution in presence of an imperfect electrical contact between the metallic strip and the waveguide walls. No resonances are observed, since the electric field is confined in the waveguide despite the existence of thin air gaps. Figure \ref{Figure4a} illustrates the simulated $|S_{21}|$ of the phase shifter for a design with and without corrugations. These corrugations, on both the metallic strip and the waveguide wall, enhance the transmission in the entire operating frequency range. When the strip corrugations are not included (blue lines), peaks in the upper part of the frequency range begin to appear due to resonances. Other deeper peaks in transmission show up in the case of not implementing the wall corrugations. Moreover, as the metallic strip employs dielectric material for its structure, additional insertion losses exist. To take into account these losses, a comparison between the phase shifter, a straight gap-waveguide and a conventional hollow waveguide is shown in Fig. \ref{Figure4b}. The increase of insertion losses because of the addition of the metallic strip is about 0.7 dB compared to the straight WR15 gap-waveguide section.

\begin{figure}[h]
	\centering
	\includegraphics[width= 0.45\textwidth]{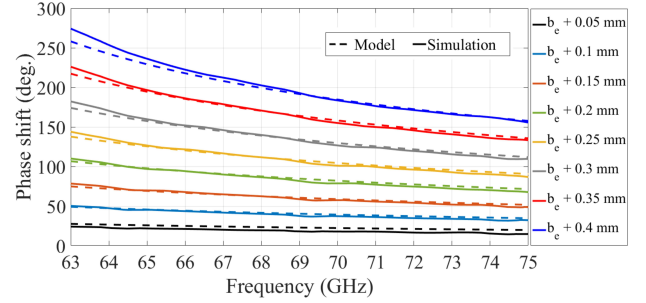}
	\caption{Simulated phase shift for different curvature radii and comparison with the proposed model for \textit{b\textsubscript{e}} = 0.55 mm.}
	\label{Figure5}
\end{figure}

The simulated phase shift results are presented in Fig. \ref{Figure5} along with the results provided by the model. Good agreement is observed between the simulation and the model results, providing more phase shift dispersion when the radius of curvature increases. For lower curvature radii, which entail larger broadside waveguide, the deviation of the phase shift tends to decrease. This is because of the resulting propagation constant, which is less dispersive when the screw does not increase the curvature radius (reference position). According to this effect, a less dispersive phase shifter can be designed by extending the phase shift area (i.e. enlarging the range of the semi-major axis \textit{a\textsubscript{e}}) and employing smaller curvature radii.

\section{Experimental Validation}

\begin{figure}[t]
	\centering
	\captionsetup{labelfont={color=black}}
	
	\subfigure[]{\includegraphics[width= 0.45\textwidth]{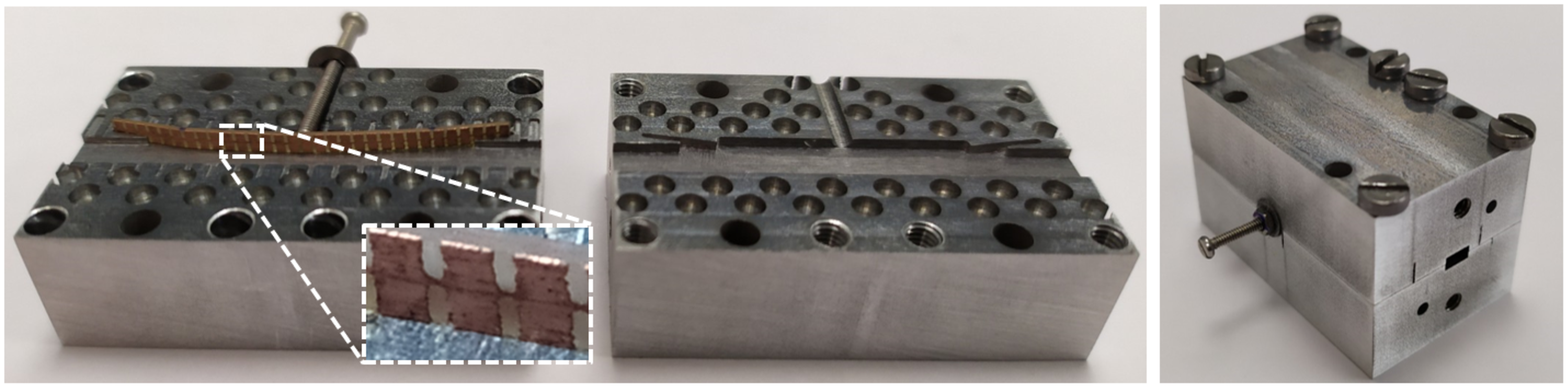}\label{Figure6a}}
	\hspace{0mm}
	\subfigure[]{\includegraphics[width= 0.4\textwidth]{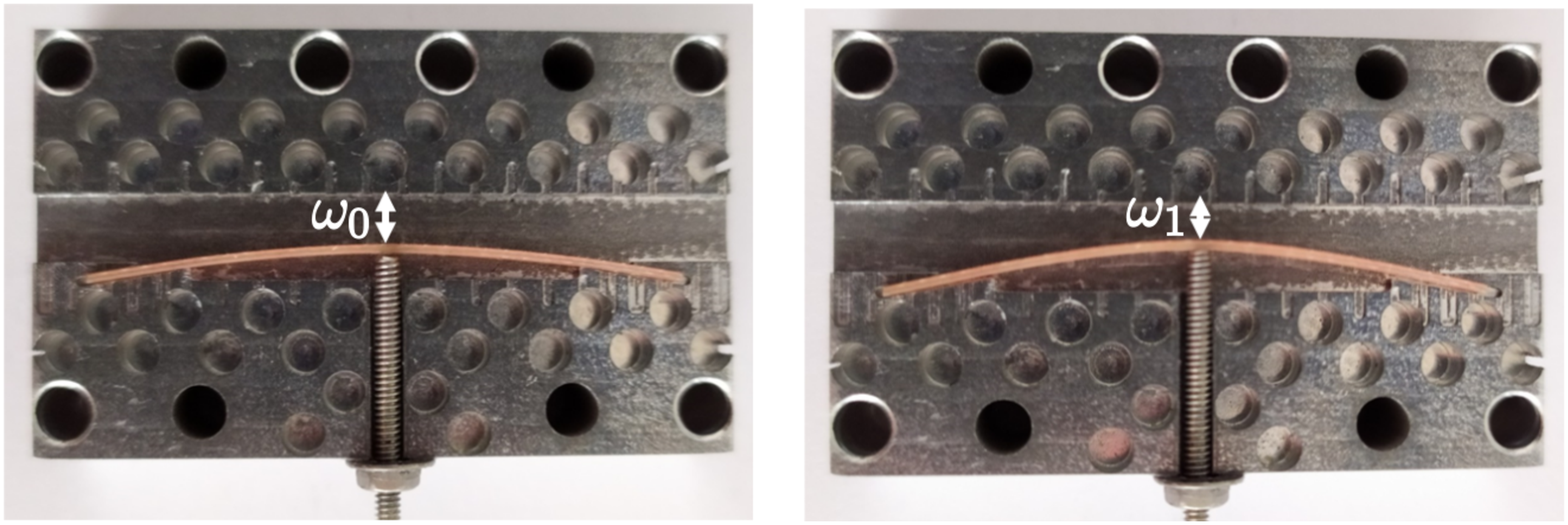}\label{Figure6b}}
	\caption{Manufactured prototype: (a) forming layers and assembled prototype, (b) detail of the modification of the curvature radius by the pressing screw.}
	\label{Figure6}
\end{figure}

\begin{figure}[t]
	\centering
	\captionsetup{labelfont={color=black}}
	
	\subfigure[]{\includegraphics[width= 0.47\textwidth]{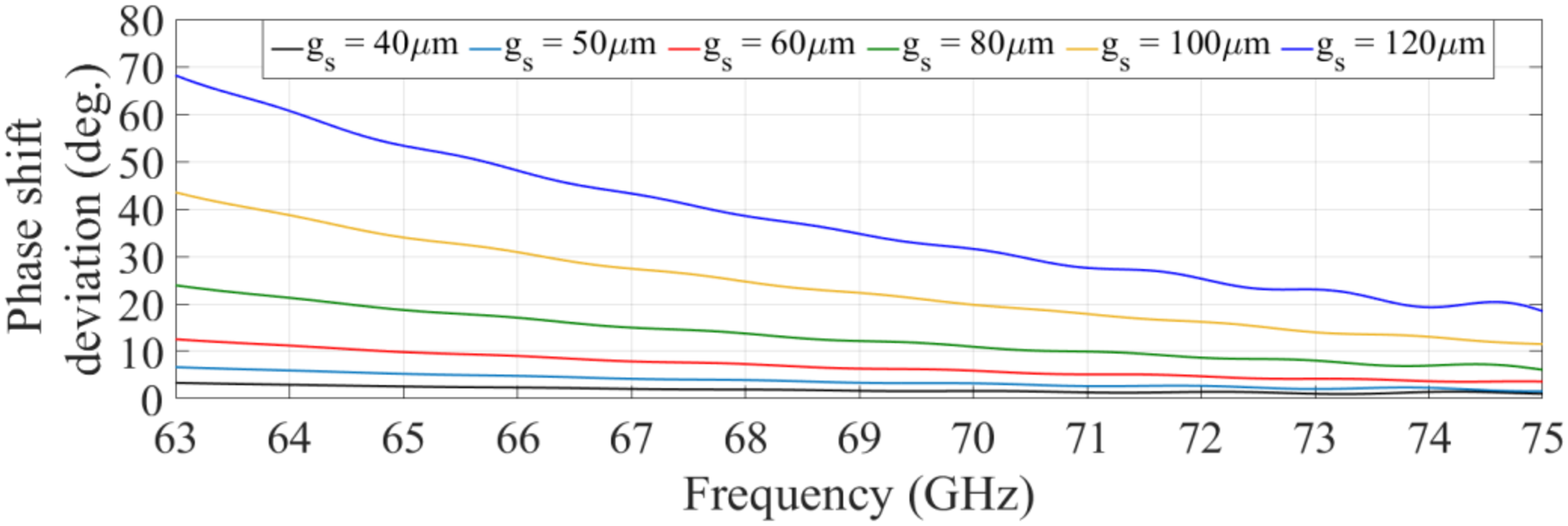}\label{Figure7a}}
	\hspace{0mm}
	\subfigure[]{\includegraphics[width= 0.47\textwidth]{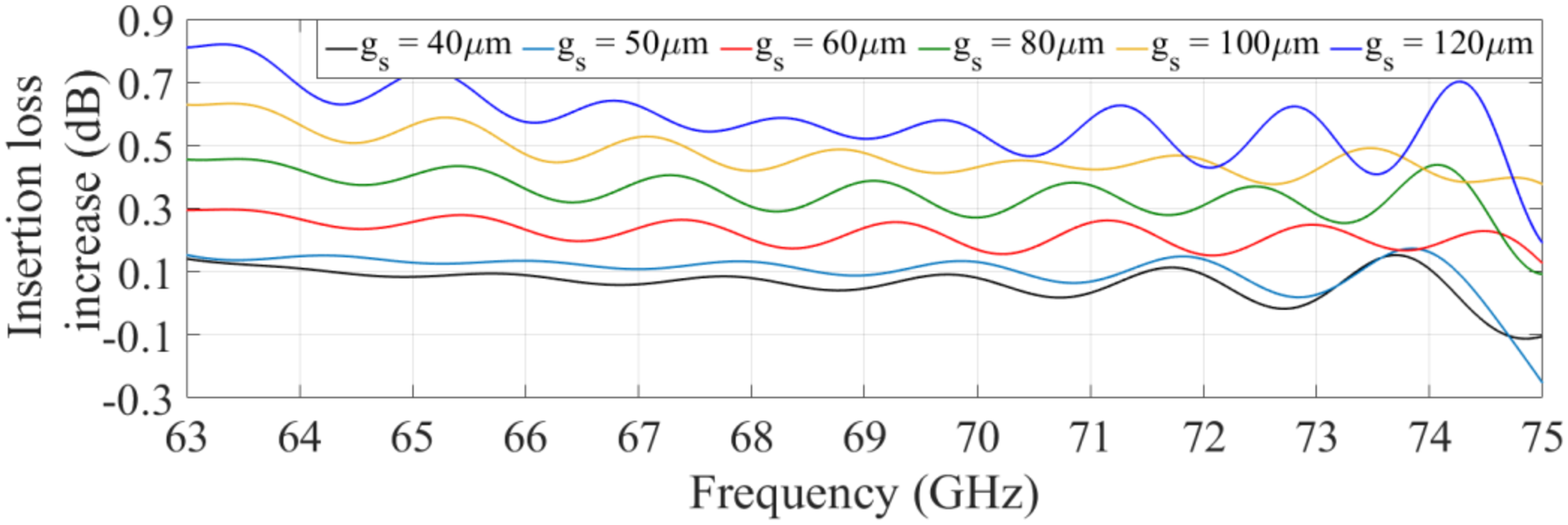}\label{Figure7b}}
	\caption{Effect of the vertical gap g\textsubscript{s} (illustrated in Fig. 3(c)) between the metallic strip and the upper and lower surfaces of the waveguide: (a) on the phase shift, and (b) on the insertion losses. The reference simulation of the phase shifter has a g\textsubscript{s} = 30 $\mu$m and a curvature radius \textit{b\textsubscript{e}} = 0.8 mm.}
	\label{Figure7}
\end{figure}

In order to validate the simulation results, a prototype of the proposed phase shifter was manufactured in CNC (Computer Numerical Control) technology. Figure \ref{Figure6a} shows the forming layers of the reconfigurable phase shifter, as well as the final assembly. Figure \ref{Figure6b} shows the effect of the pressing screw over the metallic strip. The flexible metallic strip has been manufactured in FR4 (substrate thickness of 0.4 mm and $\epsilon_r = 4.5$) to obtain the desired flexible behavior to implement the reconfigurability in the phase shifter. The gaps between the metallic strip and the lower and upper surface of the waveguide were measured to assess the correct assumption of gap sizes. The measured gaps have a size of 25$\mu$m $\pm$ 10$\mu$m. Figures \ref{Figure7a} and \ref{Figure7b} show the effect of the gap sizes in the produced phase shift and insertion losses, respectively. For gap sizes bigger than 100 $\mu$m, the gaps have a significant effect in the performance of the phase shifter because of the existence of electric field that is not guided properly by the metallic strip. Since the measured gaps are less than this value, the electric field remains confined by the metallic strip, similarly to that shown in Fig. \ref{Figure3c}.

\begin{figure}[t]
	\centering
	\captionsetup{labelfont={color=black}}

	\subfigure[]{\includegraphics[width= 0.44\textwidth]{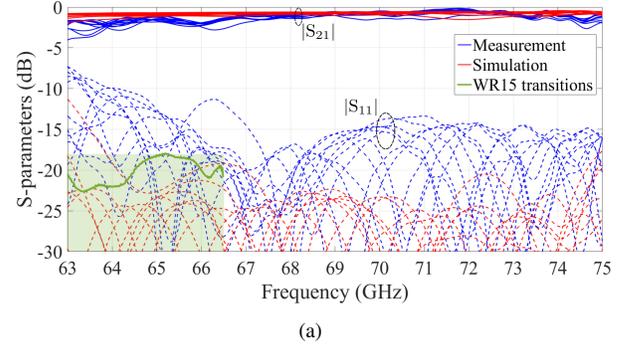}\label{Figure8a}}
	\hspace{0mm}
	\subfigure[]{\includegraphics[width= 0.35\textwidth]{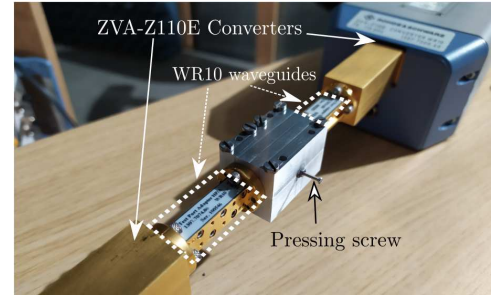}\label{Figure8b}}
	\caption{Measured S-parameters: (a) magnitude of the phase shifter for the complete range of radius of curvature and comparison with the simulated results and, (b) measurement setup in WR10.}
	\label{Figure8}
\end{figure}

The measured and simulated scattering parameters, for the complete range of curvature radius, are shown in Fig. \ref{Figure8a}. They have been measured using the ZVA-Z110E Converters that are connected to the vector network analyzer (VNA) R\&S-ZVA67. The frequency multipliers ZVA-Z110E are capable of raising the measurement range to 110 GHz because the VNA is limited approximately up to 66.5 GHz. A frequency range from 63 to 75 GHz is established. The measured results show a shift in the lowest limit frequency, achieving a $|S_{11}|$ lower than -10 dB from 64 to 75 GHz (15.8\% impedance bandwidth). This large impedance bandwidth is due to the taper in the waveguide width provided by the elliptical profile of the metallic strip. There is a smooth decrease in waveguide width up to the middle of the phase shifter and then, the waveguide width experiments a smooth increase to the width of the output port. The values obtained for the measured reflection coefficient are greater than the simulation results. This fact is due to the standard waveguide size 10 (WR10) used by the ZVA-Z110E Converters employed in the measurement setup. The transitions between WR10 and WR15 (waveguide size of the phase shifter) increase the $|S_{11}|$. The measurement setup in WR15 has not been used since it is limited up to 66.5 GHz due to the 1.85 mm coaxial to WR15 waveguide transitions. For the sake of validation, a measurement with the setup in WR15 has also been done, in the range of 63 to 66.5 GHz, and it is observed the decrease in the reflection value approaching to simulation levels (green line in Fig. \ref{Figure8a}). Furthermore, low insertion losses are obtained with a mean value of 1.7 dB and lower than 3 dB in the impedance bandwidth.

\begin{figure}[t]
	\centering
	\includegraphics[width= 0.46\textwidth]{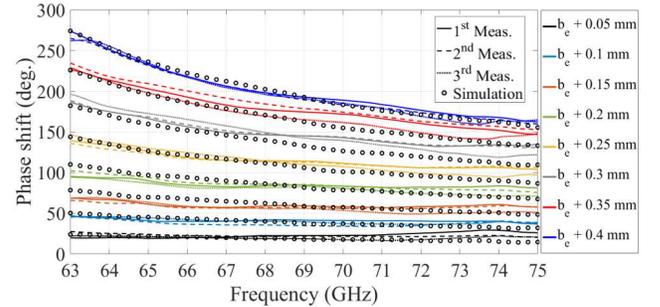}
	\caption{Measured phase shift for three consecutive measurements in comparison with the simulation results for \textit{b\textsubscript{e}} = 0.55 mm.}
	\label{Figure9}
\end{figure}

The measured phase shift is shown in Fig. \ref{Figure9}. In order to assess the repeatability of the prototype, three consecutive measurements have been carried out. This test ensures that the metallic strip returns to its original position after being deformed by the tuning screw. The screw has a standard metric 1.6 (M1.6), whose pitch thread is 0.35 mm. Therefore, an increment in b\textsubscript{ellipse} of 0.05 mm implies approximately 0.14 screw turns. The phase shift resolution will depend on the precision with which the pressing screw is turned. For instance, in the presented prototype, one turn on the pressing screw is approximately translated into a mean phase shift of 180\textsuperscript{o}. The measurement results present a good agreement with the simulation results providing the same dynamic range for the phase shift. The small discrepancies evidenced could be associated to approximating the real deformation of the metallic strip with an elliptical profile. Moreover, the repeatability of the measurements show the good performance of the flexible metallic strip to set the desired phase shift. Finally, a comparison between the proposed phase shifter and other reconfigurable phase shifters in mm-wave frequencies is illustrated in Table \ref{Table1}. The proposed reconfigurable phase shifter achieves a better performance in insertion losses and FoM (figure-of-merit) along the operating frequency\cite{LcPhaseShifter1}. The return losses (RL) achieved by our phase shifter has good performance compared with the referenced works except for \cite{Patent2}. However, that work provides low phase shift. Some of the reported tunable phase shifters require a complex setup \cite{PiezoelectricPhaseShifter} or a more complicated implementation of the tuning elements in the waveguide \cite{LcPhaseShifter3}. By applying the gap waveguide to the phase shifter design, more options in the location of the gap for the split waveguide is achieved in contrast with the traditional implementation using E-plane split waveguide \cite{LcPhaseShifter1}. In addition, compared with the commercial devices \cite{PasternackPhaseShifter, SagePhaseShifter}, the gap-waveguide phase shifter presents a reduced volume and low cost that allow a cost-effective and easy integration in a full system.

\begin{table}[h]
	\centering
	\scriptsize
	\captionsetup{justification=centering}
	\captionsetup{labelfont={color=black}}
	\setlength{\tabcolsep}{0.5pt} 
	
	\caption{\\ Comparison of reconfigurable phase shifters in mm-wave band with this work}
	\begin{tabular}{c c c c c c c c}
		\thickhline
		\textbf{Ref.} & \textbf{ \thead{ {\scriptsize Frequency} \\ {\scriptsize (GHz)}}}
		& \textbf{Technology} & \textbf{ \thead{{\scriptsize Max. IL} \\ {\scriptsize (dB)}}} & \textbf{{ \thead{{\scriptsize RL} \\ {\scriptsize (dB)}}}} &
		\textbf{ \thead{ {$\mathbf{\phi}$\textsubscript{max}}  \\ {(\textsuperscript{o})}  }}&  
		\textbf{ \thead{{FoM} \\ {(\textsuperscript{o}/dB)} }} 
		& \textbf{ \thead{ {\scriptsize Complexity/} \\ {\scriptsize Cost}}}\\ 
		
		\hline
		\cite{PiezoelectricPhaseShifter}       & 230-250         & Piezoelectric        & 3  &   10   &  380        &   127   &  High         \\ \hline
		\cite{PinDiodePhaseShifter0}       & 26-30        & PIN Diode        & 7.8    &  10  &     348.75     &  44.7    &  Low/Medium          \\ \hline
		\cite{PinDiodePhaseShifter2}       & 33-50        & PIN Diode        & 10 &   10  &     270     &  30    &  Low/Medium         \\ \hline
		\cite{LcPhaseShifter1}      &  85-110       & Liquid Crystal        & 8   &  n. a. &    900      &   125-145 &  High            \\ \hline
		\cite{LcPhaseShifter3}       & 99-105         & Liquid Crystal        &   2.7 &  6  &    318     &   118-148   &  High           \\ \hline
		\cite{Ferroelectric1}       & 30-40        & Ferroelectric       & 9     &  5  &     360     &  36-50    &  Low          \\ \hline
		\cite{Ferroelectric2}       & 50-54        & Ferroelectric       & 6  &   n. a.  &     30     &  5    &  Low          \\ \hline
		\cite{Patent2}       & \textgreater 75   &   Waveguide       &  n. a.  &   25   &    30    &  n. a.    &  Medium         \\ \hline
		\cite{PasternackPhaseShifter}, \cite{SagePhaseShifter}     & 50-75        &  Waveguide       & 1     &   18  &     180     &  180    &  High          \\ \hline
		\textbf{This work}       & 64-75         & Gap-waveguide        & 3   &    12  &  250      &   95-152    &  Low        \\ \hline
		
		\thickhline
	\end{tabular}
	\label{Table1}
\end{table}

\section{Conclusion}

This paper presents the first mechanically reconfigurable phase shifter design in gap-waveguide technology for mm-wave frequencies. The gap-waveguide technology enables a low-complex prototype that allows the inclusion of a flexible metallic strip inside the waveguide. This element provides the reconfigurability behavior of the phase shifter. A prototype has been manufactured to validate the simulation design. The experimental results show an operational frequency bandwidth from 64 to 75 GHz with averaged insertion loss of 1.7 dB and a maximum dynamic phase shift of 250\textsuperscript{o}.


\begin{thebibliography}{00}

\bibitem{MultibeamAntennas}
W. Hong \textit{et al.}, ``Multibeam Antenna Technologies for 5G Wireless Communications,'' \textit{IEEE  Trans.  Antennas  Propag.}, vol. 65, no. 12, pp. 6231-6249, Dec. 2017.

\bibitem{MechanicalPhaseShifter1}
Q. Zhang, C. Yuan and L. Liu, ``Studies on mechanical tunable waveguide phase shifters for phased-array antenna applications," in \textit{IEEE International Symposium on Phased Array Systems and Technology (PAST)}, Waltham, MA, 2016.

\bibitem{MechanicalPhaseShifter2}
L. Polo‐López, J. L. Masa‐Campos, J. A. Ruiz‐Cruz, ``Reconfigurable H‐plane waveguide phase shifters prototyping with additive manufacturing at K‐band'' \textit{Int. J. RF Microw. Comput. Aided Eng.}, vol. 29, e21980, 2019;

\bibitem{MechanicalPhaseShifter3} 
Y. Yang, C. Yuan, G. Cheng and B. Qian, ``Ku-Band Rectangular Waveguide Wide Side Dimension Adjustable Phase Shifter,'' \textit{IEEE Trans.  Plasma  Sci.}, vol. 43, no. 5, pp. 1666-1669, May 2015.


\bibitem{PiezoelectricPhaseShifter}
A. A. Ibrahim, H. N. Shaman and K. Sarabandi, ``A Sub-THz Rectangular Waveguide Phase Shifter Using Piezoelectric-Based Tunable Artificial Magnetic Conductor,'' \textit{IEEE Trans. THz Sci. Technol.}, vol. 8, no. 6, pp. 666-680, Nov. 2018


\bibitem{PinDiodePhaseShifter0}
J. G. Yang and K. Yang, ``Ka-Band 5-Bit MMIC Phase Shifter Using InGaAs PIN Switching Diodes,'' in \textit{IEEE Microw. Wireless Compon. Lett.}, vol. 21, no. 3, pp. 151-153, March 2011.

\bibitem{PinDiodePhaseShifter1}
B. Muneer, Z. Qi and X. Shanjia, ``A Broadband Tunable Multilayer Substrate Integrated Waveguide Phase Shifter,'' \textit{IEEE Microw. Wireless Compon. Lett.}, vol. 25, no. 4, pp. 220-222, April 2015.

\bibitem{PinDiodePhaseShifter2}
E. Villa, B. Aja, J. Cagigas, E. Artal and L. de la Fuente, ``Four-State Full Q-Band Phase Shifter Using Smooth-Ridged Waveguides,'' \textit{IEEE Microw. Wireless Compon. Lett.}, vol. 27, no. 11, pp. 995-997, Nov. 2017.

\bibitem{PinDiodePhaseShifter1_2}
W. J. Liu, S. Y. Zheng, Y. M. Pan, Y. X. Li and Y. L. Long, ``A Wideband Tunable Reflection-Type Phase Shifter With Wide Relative Phase Shift,'' \textit{IEEE  Trans.  Circuits  Syst.  II,  Exp.  Briefs}, vol. 64, no. 12, pp. 1442-1446, Dec. 2017.



\bibitem{PinDiodePhaseShifter3}
M. T. ElKhorassani \textit{et al.}, ``Electronically Controllable Phase Shifter with Progressive Impedance Transformation at K Band,'' \textit{Appl. Sci.}, vol. 9, no. 23, 2019.

\bibitem{PinDiodePhaseShifter3_2}
A. Singh and M. K. Mandal, ``Electronically Tunable Reflection Type Phase Shifters,'' \textit{IEEE  Trans.  Circuits  Syst.  II,  Exp.  Briefs}, vol. 67, no. 3, pp. 425-429, March 2020.

\bibitem{PinDiodePhaseShifter4}
Y. Xiong, X. Zeng and J. Li, ``A Tunable Concurrent Dual-Band Phase Shifter MMIC for Beam Steering Applications,'' \textit{IEEE  Trans.  Circuits  Syst.  II,  Exp.  Briefs}, 2020.

 \bibitem{LcPhaseShifter1}
R. Reese \textit{et al.}, ``Liquid Crystal Based Dielectric Waveguide Phase Shifters for Phased Arrays at W-Band,'' \textit{IEEE Access}, vol. 7, pp. 127032-127041, 2019.

\bibitem{LcPhaseShifter2}
A. Franc, O. H. Karabey, G. Rehder, E. Pistono, R. Jakoby and P. Ferrari, ``Compact and Broadband Millimeter-Wave Electrically Tunable Phase Shifter Combining Slow-Wave Effect With Liquid Crystal Technology,'' \textit{IEEE Trans. Microw. Theory Techn.}, vol. 61, no. 11, pp. 3905-3915, Nov. 2013.

\bibitem{LcPhaseShifter3}
M. Jost \textit{et al.}, ``Liquid crystal based low-loss phase shifter for W-band frequencies,'' \textit{Electron. Lett.}, vol. 49, no. 23, pp. 1460-1462, 7 Nov. 2013.

\bibitem{LcPhaseShifter4}
M. Jost \textit{et al.}, ``Miniaturized Liquid Crystal Slow Wave Phase Shifter Based on Nanowire Filled Membranes,'' \textit{IEEE Microw. Wireless Compon. Lett.}, vol. 28, no. 8, pp. 681-683, Aug. 2018.

\bibitem{LcPhaseShifter5}
A. Alex-Amor, \textit{et al.}, ``Generalized Director Approach for Liquid-Crystal-Based Reconfigurable RF Devices,'' \textit{IEEE Microw. Wireless Compon. Lett.}, vol. 29, no. 10, pp. 634-637, Oct. 2019.

\bibitem{Ferroelectric1}
Z. Zhao, X. Wang, K. Choi, C. Lugo and A. T. Hunt, ``Ferroelectric Phase Shifters at 20 and 30 GHz,'' \textit{IEEE Trans. Microw. Theory Techn.}, vol. 55, no. 2, pp. 430-437, Feb. 2007.

\bibitem{Ferroelectric2}
Z. Wang \textit{et al.}, ``Millimeter wave phase shifter based on ferromagnetic resonance in a hexagonal barium ferrite thin film'', \textit{Appl. Phys. Lett.}, vol. 97, no. 7, Aug. 2010.

\bibitem{Ferroelectric3}
V. Sharmaa, Y. Khivintsev, I.Harward, B.J. Kuanr and Z. Celinski, ``Fabrication and characterization of microwave phase shifter in microstripgeometry with Fefilm as the frequency tuning element'' \textit{J. Magn. Magn. Mater.}, vol. 489, no. 165412, June 2019.

\bibitem{GapWave1}
P. Kildal, E. Alfonso, A. Valero-Nogueira and E. Rajo-Iglesias, ``Local Metamaterial-Based Waveguides in Gaps Between Parallel Metal Plates,'' \textit{IEEE Antennas  Wirel.Propag. Lett.}, vol. 8, pp. 84-87, 2009.

\bibitem{GapWavePhaseShifter1} 
E. Rajo-Iglesias, M. Ebrahimpouri and O. Quevedo-Teruel, ``Wideband Phase Shifter in Groove Gap Waveguide Technology Implemented With Glide-Symmetric Holey EBG,'' \textit{IEEE Microw. Wireless Compon. Lett.}, vol. 28, no. 6, pp. 476-478, June 2018.

\bibitem{GapWavePhaseShifter2}
\'{A}. Palomares-Caballero, A. Alex-Amor, P. Padilla, F. Luna and J. Valenzuela-Valdes, ``Compact and Low-Loss V-Band Waveguide Phase Shifter Based on Glide-Symmetric Pin Configuration,'' \textit{IEEE Access}, vol. 7, pp. 31297-31304, 2019.

\bibitem{GapWave2} 
M. Ebrahimpouri, E. Rajo-Iglesias, Z. Sipus and O. Quevedo-Teruel, ``Cost-Effective Gap Waveguide Technology Based on Glide-Symmetric Holey EBG Structures,'' \textit{IEEE Trans. Microw. Theory Techn.}, vol. 66, no. 2, pp. 927-934, Feb. 2018.


\bibitem{Patent1}
J. Reindel, ``Variable Printed Circuit Waveguide Filter,'' U.S. Patent 4 990 871, Feb. 5, 1991.

\bibitem{Patent2}
K. W. Brown, D. M. Gritters and A. W. Chang, ``Waveguide Mechanical Phase Adjuster,'' U.S. Patent 9 196 940, Nov. 24, 2015.

\bibitem{PasternackPhaseShifter}
Pasternack, ``0 to 180 Degree WR-15 Waveguide Phase Shifter'', 2018. Available: \url{https://www.pasternack.com/images/ProductPDF/PE-W15PS1001.pdf}

\bibitem{SagePhaseShifter}
SAGE Millimeter, Inc., ``V-band Micrometer Driven Phase Shifter'', 2015. Available: \url{https://sftp.eravant.com/content/datasheets/STP-18-15-M2.pdf}

\bibitem{Pozar}
D. M. Pozar. \textit{Microwave Engineering}, 4th edition, Wiley, 2011.



\bibitem{Corrugations}
A. Vosoogh, H. Zirath and Z. S. He, ``Novel Air-Filled Waveguide Transmission Line Based on Multilayer Thin Metal Plates,'' \textit{IEEE Trans. THz Sci. Technol.}, vol. 9, no. 3, pp. 282-290, May 2019.

\bibitem{Corrugations2}
A. Tamayo-Dom\'inguez, H. Azkiou, J. M. Fern\'andez-Gonz\'alez and O. Quevedo-Teruel, ``Space Reduction Between Parallel Gap Waveguides Using Stacked Glide-Symmetric Metal Sheets,'' in \textit{13th European Conference on Antennas and Propagation (EuCAP)}, Krakow, Poland, 2019.







\end{thebibliography}
\end{document}